\documentclass[11pt]{article}

\usepackage{color}
\usepackage{hyperref}

\hypersetup{
    colorlinks=true,
    linkcolor=blue,
    filecolor=red,
    urlcolor=magenta,
    breaklinks=true,
}

\usepackage{color}
\usepackage{amsmath}
\usepackage{amssymb}
\usepackage{multirow, varwidth}
\usepackage{epsfig}
\usepackage{verbatim}
\makeatletter
\newif\if@restonecol
\makeatother

\usepackage{xr}

\DeclareRobustCommand\onedot{\futurelet\@let@token\@onedot}
\def\onedot{.} %
\def\eg{\emph{e.g}\onedot} 
\def\ie{\emph{i.e}\onedot}

\def\etal{\emph{et al}\onedot}

\usepackage{color,booktabs, graphicx, fullpage}
\usepackage[square,numbers]{natbib}
\usepackage{indentfirst}

\begin{document}

\title{\huge On the Virality of Animated GIFs on Tumblr}
\date{}
\author{
Yunseok Jang\thanks{This work has been done when Yunseok and Yale were at Yahoo Research, New York, NY.} \\
University of Michigan \\
Ann Arbor, MI
\and
Yale Song$^*$ \\
Microsoft Research \\
Redmond, WA
\and
Gunhee Kim \\
Seoul National University \\
Seoul, Korea
}
\maketitle

\begin{abstract}
Animated GIFs are becoming increasingly popular in online communication. People use them to express emotion, share their interests and enhance (or even replace) short-form texting; they are a new means to tell visual stories. Some creative animated GIFs are highly addictive to watch, and eventually become \textit{viral} -- they circulate rapidly and widely within the network. What makes certain animated GIFs go viral? In this paper, we study the virality of animated GIFs by analyzing over 10 months of \textit{complete} data logs (more than 1B posts and 12B reblogs) on \textit{Tumblr}, one of the largest repositories of animated GIFs on the Internet. We conduct a series of quantitative and comparative studies on Tumblr data, comparing major types of online content -- text, images, videos, and animated GIFs. We report on a number of interesting, new findings on animated GIFs. We show that people tend to make animated GIFs easily searchable and discoverable by adding more hashtags than other content types. We also show that animated GIFs tend to go more viral than images and videos on Tumblr. With more in-depth analysis, we present that animated GIFs tend to get reblogged more and followed more from non-followers, while animated GIFs have more recurrence of a post. Lastly, we show that the virality of animated GIFs is more easily predictable than that of images and videos.
Our findings have implications for characterizing the use of content types over social media services; it could also bring important impacts on social media services including marketing, personalization, and recommendation.
\end{abstract}

\section{Introduction}
\label{sec:introduction}

Animated GIFs have had several ups and downs throughout the evolution of social networks~\cite{eppink-jcv-14}. In the early 90's, GIFs were used in personal websites to depict rudimentary, untextured shapes, such as American flags and ``under construction'' signs. In the late 90's, GIFs lost their popularity to emojis used in AOL Instant Messenger. In the mid 2000's, MySpace led a GIF resurgence with personal pages, but its cluttered user interface caused losing its throne to GIF-free Facebook in the late 2000's; GIF was gone again. In the early 2010's, Tumblr and Reddit dramatically increased the popularity of GIFs again. Today, Tumblr remains as one of the \textit{de facto} go-to places for GIF enthusiasts.

The increasing popularity of GIFs inspired people to find new, creative ways to consume the content, such as memes, cinemagraphs, photojournalism, advertisement, video highlighting and summarization. Some GIFs are noticed quickly by people and become an Internet sensation; in other words, they go \textit{viral} -- they circulate rapidly and widely across social media platforms. Reflecting this, various press and media published articles about the virality of animated GIFs~\cite{moreau-press-15,johnson-press-15,barrett-press-16}. While good at providing timely information to the general audience, those articles typically provide anecdotal evidence and subjective opinions to support their claims.

In terms of scholarly work, there have been some efforts to understand the virality of traditional content types, such as short texts (tweets)~\cite{hansen-fit-11,guerini-icwsm-11,goel-ms-15}, images~\cite{guerini-sc-13,deza-cvpr-15}, and videos~\cite{jiang-icmr-14}. Also, there has been some recent work on animated GIFs regarding sentiment analysis~\cite{jou-mm-14}, engagement analysis~\cite{bakhshi-chi-16}, captioning~\cite{li-cvpr-16}, and generation~\cite{gygli-cvpr-16}. However, to the best of our knowledge, no prior work exists that focuses on studying the virality of animated GIFs.

In this paper, we present a large-scale analysis of GIF virality on a popular social network, Tumblr -- a microblogging platform with 336 million blogs and 146 billion posts.\footnote{\url{https://www.tumblr.com/about} (accessed Feb 12, 2017)} We focus on the following aspects of GIF virality: How does GIF virality compare to that of images and videos? Do GIFs tend to go more viral than images and videos? What makes certain animated GIFs go viral? Can a machine automatically predict the virality of animated GIFs?

To answer these questions, we conduct several quantitative and comparative studies on GIF virality by analyzing over 10 months of \textit{complete} data on Tumblr, amounting to more than 1 billion posts and 12 billion reblogs. We start by investigating user activities around different content types on Tumblr (Section~\ref{sec:tumblr_data}). We then investigate various statistical patterns related to the virality of visual content, comparing GIFs against images and videos (Section~\ref{sec:virality}). Next, we study what makes certain animated GIFs go viral by performing correlation analyses of different features extracted from textual, visual, and social cues (Section~\ref{sec:prediction_by_content}). Finally, we provide empirical evidence showing GIF virality is easier to predict than that of images and videos.

Our work provides the first-of-its-kind, comprehensive, data-driven, statistical evidence around multiple aspects of GIF virality. Our major findings include:
\begin{itemize}

\item To make GIF posts easily searchable and discoverable in Tumblr, people use 47.59\% shorter captions and 29.18\% more hashtags on GIF posts than other posts.%

\item Animated GIFs spread deeper by 23.62\% and wider by 208.27\% than images and videos, suggesting they tend to go more viral than the others. They also have a 10.47\% higher virality score than images and videos (measured by the Wiener Index~\cite{goel-ms-15}). %

\item Comparing with images and videos, animated GIFs are reblogged 174.83\% more from the non-followers, and the non-followers are 15.56\% more likely to initiate following the post owners after reblogging. %
It shows that including animated GIF on its post makes it more likely to spread out beyond the existing follower-followee network, and at the same time making the existing network richer.

\item Animated GIFs have 31.97\% more number of recurrence (\ie~resurface in popularity multiple times) with 3.24\% shorter lull period in between the first and its subsequent peaks (\ie~local maximums of daily reblog count) than images and videos. Moreover, animated GIFs lead that the second peaks occur 19.42\% more, and the first peaks continue 5.52\% longer.

\item GIF virality is 4.47\% easier to predict than that of images and videos. Unlike previous findings that showed social features are most predictive of image virality~\cite{totti-websci-14,deza-cvpr-15}, we show that textual features are most predictive of GIF virality.

\end{itemize}

\section{Related work}
\label{sec:related_work}
\smallskip
\textbf{Tumblr}.
Tumblr has gained much attention recently from the data mining community. Chang~\etal~\cite{chang-kddexpr-14} conduct one of few early studies on Tumblr and provide a comprehensive statistical overview of Tumblr, comparing it with Twitter and Facebook. The work of Xu~\etal~\cite{jiejun-websci-14} is another quantitative research on behavioral patterns on Tumblr.

Vahabi~\etal~\cite{vahabi-cikm-15} propose a recommender system for Tumblr. Their idea is to design a mathematical model of recommendation by incorporating information diffusion over social networks, such as liking, following, reblogging. Shin~\etal~\cite{shin-cikm-15} also address recommendation on Tumblr, but develop a different approach, \textit{boosted inductive matrix completion} (BIMC). The BIMC consists of two components: one for the low-rank structure of a Tumblr follower graph, another for the latent structure of side-information of users and blogs. Grbovic~\etal~\cite{grbovic-kdd-15} present a framework for sponsored post advertising on Tumblr, especially aiming at gender and interest targeting advertisement.%

\smallskip
\textbf{Animated GIFs}. Although animated GIFs have emerged as a new interesting form of visual communication, scholarly work on animated GIFs are still few, and their usage patterns are largely under-explored. Jou~\etal~\cite{jou-mm-14} develop a prediction model of perceived emotion for animated GIFs. Gygli~\etal~\cite{gygli-cvpr-16} present a method to automatically create GIFs from video by detecting highlights. Li~\etal~\cite{li-cvpr-16} suggest a new dataset for automatic GIF captioning task.

The work of Bakhshi~\etal~\cite{bakhshi-chi-16} is one of the closest to our work, in which they study what makes an animated GIF \textit{engaging} by analyzing 3.9 million posts and conducting interviews with 13 users on Tumblr. They show that engagement patterns of GIFs are attributed to the lack of sound, the immediacy of the communication, the availability of motion with minimum bandwidth and time, and the ability to tell a story. Although they compare like and reblog distributions according to data types (\ie~text, videos, GIFs, JPGs, and PNGs), their analysis and results are largely based on qualitative user studies done via interviews with 13 users.

\smallskip
\textbf{Virality analysis of visual content}.
Recent work used computer vision techniques to discover the relationship between image content and user behavior, measuring the popularity and virality of images over various social network platforms. Hu~\etal~\cite{yuheng-icwsm-14} analyze user photos on Instagram. Denton~\etal~\cite{emily-kdd-15} propose an approach for predicting hashtags using both image content and user information. Khosla~\etal~\cite{khosla-www-14} measure image popularity as the number of views over Flickr images, and predict image popularity using visual and social attributes.

Guerini~\etal~\cite{guerini-sc-13} study image virality on Google Plus; they measure the virality using the magnitude of user activities (\eg +1 votes, replies and reshares) and analyze correlation with visual features. Deza~\etal~\cite{deza-cvpr-15} study image virality on \textit{reddit} using social interaction patterns, measuring virality using quantitative metadata. Totti~\etal~\cite{totti-websci-14} conduct a similar study on Pinterest, to predict image popularity as the number of repins, using various visual, aesthetic, semantic, and social features of the image.

Compared to all previous studies, our work is unique in that we conduct quantitative and comparative studies about the virality of three types of visual content: static photos, animated GIFs, and videos. Specifically, the following analyses have not been reported yet with a particular focus on animated GIFs: (i) comparison of statistics of reblog cascades and structural virality measures; (ii) discovery of the relations between the virality and post content features.

\begin{figure}[t]
  \centering
  \includegraphics[width=0.8\textwidth]{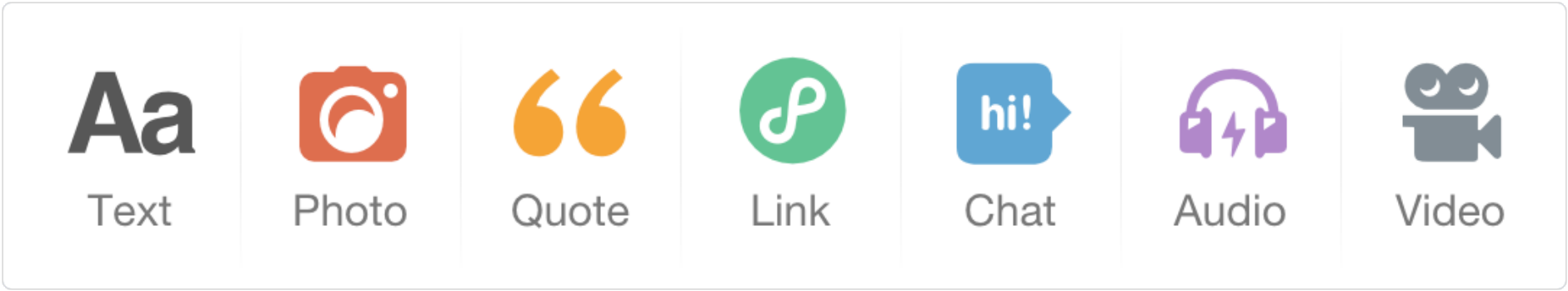}
    \vspace{-20pt}
  \caption{Seven post types supported on Tumblr.}
  \label{fig:post_types}
\end{figure}

\section{Tumblr}
\label{sec:tumblr_data}

\begin{figure}[t]
  \begin{center}
  \includegraphics[width=.95\textwidth]{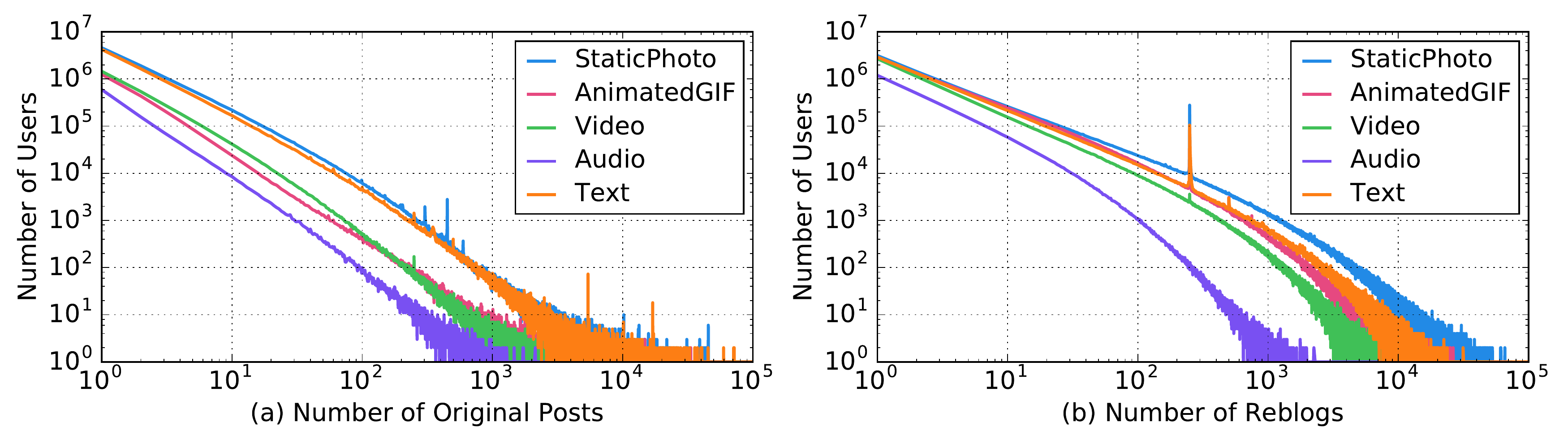}
    \end{center}
    \vspace{-20pt}
    \caption{Distributions of posts per user on each post type (log scale). (a) Number of original posts per user. (b) Number of reblogged posts per user over the ten months of our study.}
  \label{fig:posts_per_user}
\end{figure}

Tumblr is a microblogging platform where users interact with short-form blog posts with multimedia content. Users create a blog post by choosing one of the seven post types shown in Figure~\ref{fig:post_types}; photo posts include both static images and animated GIFs. Users can annotate their posts with hashtags to help make content easily searchable and discoverable. Additional metadata include titles and captions.

\subsection{Data Collection}
\label{sec:data_collection}

We run our study from March to December 2016, during which we crawled the entire user activities on Tumblr. To the best of our knowledge, this is the largest study to date conducted on Tumblr data.

We note that Chang~\etal~\cite{chang-kddexpr-14} report slightly different statistics and show that \textit{Photo} is the most dominant type (78.11\%); that work was performed two years before ours, counted both original and reblogged posts, and did not distinguish static photos from animated GIFs. Also, our dataset is about 23 times larger than theirs. %

Tumblr users post a new content either by creating an original one or by reblogging an already existing one. As we will explain later, reblogging activities play an important role in our virality analysis. We, therefore, treat original posts that are created only during the first seven months of our study, from March to September 2016, so that we have enough reblogged activities for each post. As a result, we study 728 million original posts. Reblog activities of the original posts are collected over the ten months' period.

For the purpose of our study, we focus on five major content types: \textit{Text, StaticPhoto, AnimatedGIF, Audio}, and \textit{Video}. We group four post types (Text, Quote, Link, Chat) into \textit{Text}. We split photo posts into \textit{StaticPhoto} and \textit{AnimatedGIF} by checking the image file format.\footnote{Although not all GIFs are animated GIFs, we checked that over 99\% of GIFs are animated GIFs by counting the number of frames in each GIF file.} A photo post can have multiple images; we consider multi-image posts \textit{AnimatedGIF} as long as one of them is an animated GIF. All other photo posts are considered to be \textit{StaticPhoto}.

\subsection{Data Analysis}
\label{sec:data_analysis}

\smallskip
\textbf{User activity}.
We compare user activities on original posts and reblogged posts. Figure~\ref{fig:posts_per_user} shows that creation and reblogging trends can be clustered into three groups: \{\textit{AnimatedGIF, Video}\}, \{\textit{StaticPhoto, Text}\}, and \{\textit{Audio}\}. We make an interesting observation: Even though animated GIFs are considered to be a \textit{Photo} post on Tumblr, its consumption pattern is much more similar to videos rather than static photos -- this is not surprising given that an animated GIF is merely a looping short video without sound. This suggests a possible change to Tumblr user interface: GIF posts could either be categorized into \textit{Video} or as its own.\footnote{We note that, the Tumblr mobile app added a new post type dedicated to animated GIFs on July 21, 2016. The desktop version keeps the seven post types shown in Figure~\ref{fig:post_types}. See \url{http://unwrapping.tumblr.com/post/147771981997/version-6-5-tumblr-app}}

\begin{figure}[t]
  \centering
  \includegraphics[width=0.93\textwidth]{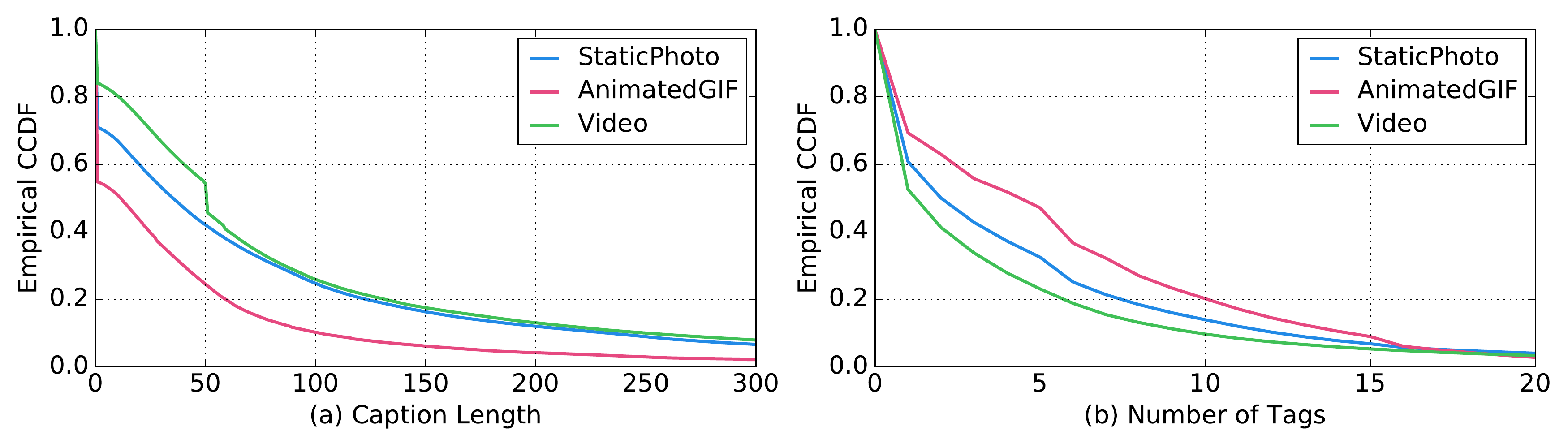}
   \vspace{-10pt}
  \caption{People use captions and hashtags differently depending on the content type. Shown here are CCDFs of (a) the caption length and (b) the number of tags. The results show that animated GIF posts have shorter captions and more hashtags compared to static photo and video posts.}
  \label{fig:visual_content_distribution}
\end{figure}

\begin{figure}[t]
  \centering
  \includegraphics[width=0.93\textwidth]{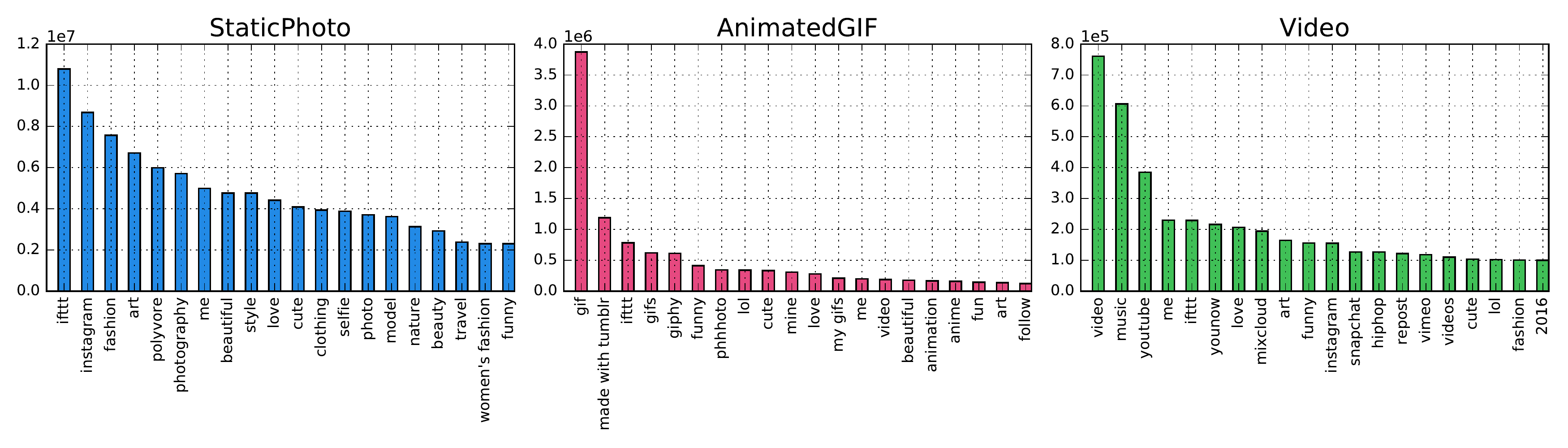}
   \vspace{-10pt}
  \caption{20 most frequent hashtags used in \textit{StaticPhoto}, \textit{AnimatedGIF}, and \textit{Video} posts.}
  \label{fig:frequent_tags}
\end{figure}

\smallskip
\textbf{Captions and hashtags}.
Previous research has shown that captions and hashtags play an important role in delivering the main concept of visual content~\cite{chang-kddexpr-14}. Here, we go one step further and show that people use captions and hashtags differently depending on the content type. We do this by analyzing captions and hashtags associated with visual posts, in terms of caption lengths, hashtag counts, and distributions.
We pre-process each hashtag into lower case and include the hashtags that contain only ASCII characters and are searchable via Tumblr API.\footnote{\url{https://www.tumblr.com/docs/en/api/v2}.}

Figure~\ref{fig:visual_content_distribution} shows complementary cumulative distribution functions (CCDFs) of (a) the length of captions (measured by character counts, excluding HTML tags) and (b) the number of hashtags. It shows the trend of GIF posts having shorter captions with more hashtags: On average, GIF posts contain 47.59\% shorter captions and 29.18\% more hashtags compared to image and video posts (see Table~\ref{tab:caption_and_hashtag_stat}). The two-sample Kolmogorov-Smirnov (KS) tests with 10,000 random samples from each group show that the differences are statistically significant at $p<10^{-96}$.

\begin{table}[t]%
  \small
  \centering
    \begin{tabular}{|l|c|c|}
    \hline
    \bf \hfill Content type \hfill  & \bf Length  & \bf \#Tags \\\hline
    \textit{StaticPhoto}            & 91.63       & 3.98 \\
    \textit{AnimatedGIF}            & \bf 48.02   & \bf 5.14  \\
    \textit{Video}                  & 108.52      & 3.09  \\\hline
    \end{tabular}
    \caption{Mean caption length and number of hashtags on three types of visual content.\label{tab:caption_and_hashtag_stat}}
\end{table}

We find one possible reason for animated GIF posts having shorter captions than other content types: people use terse sentences to make them humorous, whose effect could be diminished with verbose text (c.f., Internet memes).
Also, adding more hashtags helps the GIF posts easily searchable and discoverable, which makes them potentially become more viral.  To see this quantitatively, we analyze whether the number of hashtags in GIF posts has a positive correlation with the structural virality, and compare with those of photos and videos. Pearson correlation coefficient between the number of hashtags and the structural virality, measured by the Wiener index (explained in detail in next section), was higher in \textit{AnimatedGIF} (0.1028) then \textit{Video} (0.0499) and \textit{StaticPhoto} (0.0130).

Figure~\ref{fig:frequent_tags} shows 20 most frequent hashtags used in three types of visual content. Each content type has different distribution of popular hashtags. For example, from photo posts, we see more occurrences of \textit{fashion, art, } than others, while from GIF posts we see more occurrences of \textit{lol, mine, funny} than others. %
We observe that animated GIF and video posts often contain hashtags that represent their content types: ``gif'' and ``video,'' respectively; this rarely happens in photo posts. We believe this reflects the skewed distribution of the three content types; users may want their animated GIFs and videos to be easily searchable and discoverable by explicitly marking them with the content type.

\begin{figure}[t]
  \centering
  \includegraphics[width=0.93\textwidth]{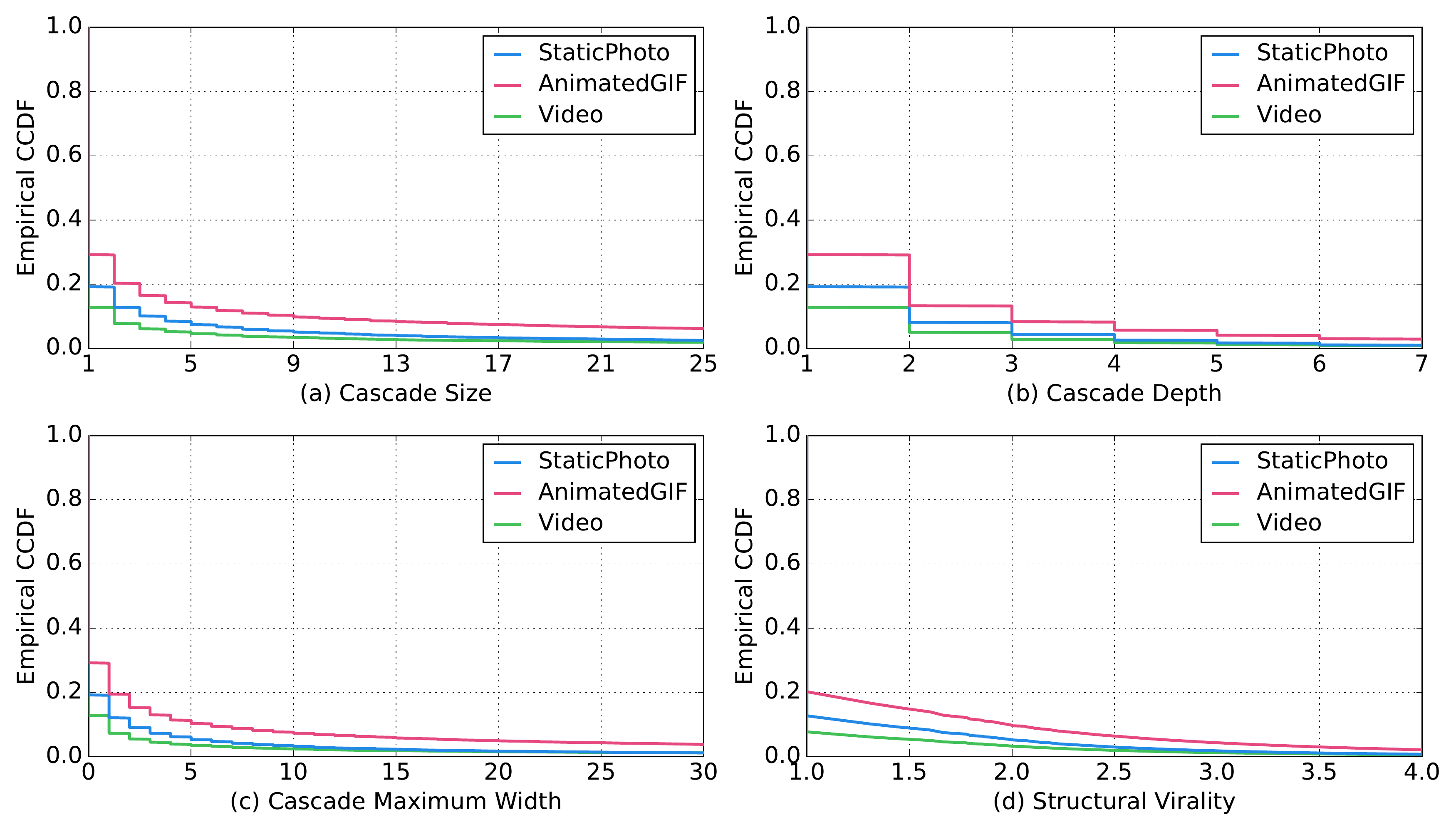}
   \vspace{-10pt}
  \caption{Virality patterns differ by content types. Shown here are CCDFs of (a) cascade size, (b) cascade depth, (c) cascade max width, and (d) structural virality (Wiener index). The results show that animated GIF posts spreads wider and deeper than static photos and videos.}
  \label{fig:cascade_info}
\end{figure}

\section{Virality Analysis}
\label{sec:virality}

We now turn to virality analysis and study whether animated GIFs have different patterns of virality than static photos and videos.

\smallskip
\textbf{Reblog cascades}.
One common approach to study virality is via cascade graphs, which model patterns of information spreading in social networks~\cite{cheng-www-14,chang-kddexpr-14,goel-ms-15}. There are two different types of cascade graphs that can be used for analyzing virality: (i) a \textit{network cascade graph} based on follower-followee relationships, and (ii) a \textit{reblog cascade tree} based on the reblog order of the content. We use the latter approach; we construct reblog cascades of 417 million original posts that include visual content in our dataset.
Formally, a reblog cascade is defined as a tree-structured graph, with a root node representing the original post, subsequent nodes representing reblog posts, and edges between nodes representing the reblog history.

Figure~\ref{fig:cascade_info} shows CCDFs of (a) cascade size, (b) cascade depth, and (c) cascade maximum width. Together they show that animated GIFs have larger, wider, and deeper cascades. Our analysis reveals that animated GIFs spread 23.62\% deeper and 208.27\% wider on average than other visual content (see Table~\ref{tab:cascade_stats}). The two-sample KS tests with 10,000 random samples show that the differences between GIFs and others are statistically significant at $p<10^{-44}$. Our results suggest strong trends for the virality of animated GIFs. Next, we quantify the virality of different visual content types.

\begin{table}[t]%
  \small
  \centering
    \begin{tabular}{|l|c|c|c|}
    \hline
    \bf \hfill Content type \hfill  & \bf Size  & \bf Width & \bf Depth \\\hline
    \textit{StaticPhoto}            & 8.27      & 2.75      & 1.40 \\
    \textit{AnimatedGIF}            & \bf 23.67 & \bf 8.71  & \bf 1.74 \\
    \textit{Video}                  & 12.94     & 2.82      & 1.28 \\\hline
    \end{tabular}
    \caption{Mean size, width and depth of reblog cascades, on three content types.\label{tab:cascade_stats}}
\end{table}

\smallskip
\textbf{Structural virality}.
Structural virality of a cascade graph is defined as a continuous quantity from being \textit{broadcast} to being \textit{viral}~\cite{goel-ms-15}. Broadcast indicates that a large number of nodes are influenced by a single root node, while viral comprises a multi-generational branching process for node infection. Simply put, structural virality is high if a cascade is deep and relatively few nodes are directly connected to the root node.

We quantify structural virality of a cascade tree $T$ using the Wiener Index~\cite{cheng-www-14,goel-ms-15},
\begin{align}
\label{eq:wiener_index}
\nu (T) = \frac{1}{n(n-1)} \sum_{i=1}^n \sum_{j=1}^n d_{ij}
\end{align}
where $n$ is the number of nodes in $T$, and $d_{ij}$ is the length of the shortest path between nodes $i$ and $j$. In words, $\nu (T)$ is the average distance between all pairs of nodes in $T$.

\begin{figure}[t]
  \centering
  \includegraphics[width=0.93\textwidth]{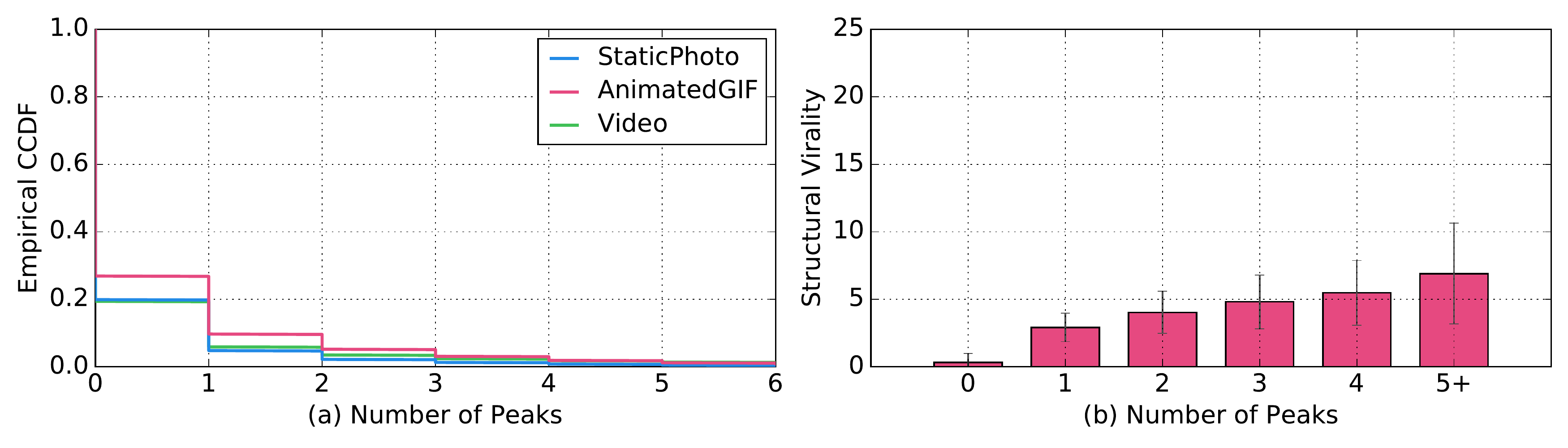}
   \vspace{-10pt}
  \caption{(a) CCDFs of the number of peaks by content types, and (b) correlation between the number of peaks and the structural virality for animated GIF posts. }
  \label{fig:peak_distribution}
\end{figure}
\begin{table}[t]%
  \small
  \centering
    \begin{tabular}{|l|c|c|c|c|}
    \hline
    \bf \hfill Content type \hfill  & \bf \# Peaks  & \bf Duration of 1st Burst & \bf Days btw. 1st/2nd Peak & \bf \% Recurrence \\\hline
    \textit{StaticPhoto}            & 0.3023 & 1.70 & 29.04 & 23.99\%     \\
    \textit{AnimatedGIF}            & \bf 0.4953 & \bf 1.91 & \bf 28.10 & \bf 35.79\%   \\
    \textit{Video}                  & 0.3753 & 1.81 & 33.38 & 29.97\%     \\\hline
    \end{tabular}
    \caption{Statistics related with recurrence on three content types.\label{tab:recurrence_stats}}
\end{table}

Figure~\ref{fig:cascade_info} (d) shows that animated GIFs have a higher structural virality ($\mu=2.02$, $\sigma=1.49$) than static photos ($\mu=1.82$, $\sigma=1.29$) and videos ($\mu=1.83$, $\sigma=1.54$) on average. The two-sample KS tests with 10,000 random samples from each group show that the differences are statistically significant at $p<10^{-44}$. Since posts with no reblogs cannot be measured with the Wiener index, we consider them as zero.

\smallskip
\textbf{What makes a post go viral}?
In the previous analysis, we observed that posts with animated GIFs have a higher structural virality, with deeper and wider reblog cascades. However, it does not explain what makes such phenomenon.

To find an answer, we calculate the information about the \textit{peaks} and the \textit{bursts} for our dataset of 417 million cascades.
Among various methods to compute \textit{peaks} and \textit{bursts}, we adopt the approach explained in Cheng~\etal~\cite{cheng-www-16}.
For each cascade, we sort its reblogs in a chronological order, discretize them by day, and plot the number of reblogs on each day.
The \textit{peaks} are detected by finding locally maximum points in the graph, and the post is called to \textit{recur} if two or more peaks are found.
The \textit{burst} indicates the duration of days for each peak, by calculating the number of days monotonically increasing and decreasing around each peak.
Cheng~\etal~\cite{cheng-www-16} introduce four variables for parameterizing the recurrence. First, each peak must have a height greater than or equal to $h_0$, a threshold for the number of reblogs in a day, and $m$ times higher than the mean reblogs per day. Also, the maximum value before and after the peak should be maintained at least $w$ days. Finally, between any two adjacent peaks, the number of reblogs should drop below $v$ times the height of the lower peak between the two, to make sure that the two adjacent peaks are distinctive enough.
We set the parameters to $h_0 = 7, w = 5, m=2, v=0.5$.

\begin{table}[t]%
  \small
  \centering
    \begin{tabular}{|l|c|c|}
    \hline
    \bf \hfill Content type \hfill  & \bf \#non-follower  & \bf \#after-follow \\\hline
    \textit{StaticPhoto}            & 2.90      & 0.45 \\
    \textit{AnimatedGIF}            & \bf 7.97  & \bf 0.52 \\
    \textit{Video}                  & 2.31      & 0.18 \\\hline
    \end{tabular}
    \caption{Mean number of non-followers and after-followers that reblog a post.\label{tab:out_network_reblogger_stats}} %
\end{table}

Table~\ref{tab:recurrence_stats} and Figure~\ref{fig:peak_distribution} (a) reveal that animated GIFs have 31.97\% more number of recurrence, and the first peaks continue 5.52\% longer. They also illustrate that animated GIFs lead the second peaks occur 19.42\% more with 3.24\% shorter lull period in between the first and its subsequent peaks than images and videos. The two-sample KS tests with 10,000 random samples from each group show that the differences between GIFs and others are statistically significant at $p<10^{-11}$ for the number of peaks, $p<10^{-14}$ for the duration and the first burst, and $p<0.006$ for the days between the first and the second peak.

Figure~\ref{fig:peak_distribution} (b) shows the correlation between the number of peaks and the structural virality. For all content types, the Pearson coefficients of correlation between the structural virality and the number of peaks are very high: \textit{StaticPhoto}: 0.7184, \textit{AnimatedGIF}: 0.7475, \textit{Video}: 0.6530, among which the \textit{AnimatedGIF} is the highest.
It concludes that the number of peaks has a strong positive correlation with the structural virality, which is more dominant in animated GIFs than other content types.

\begin{figure}[t]
  \centering
  \includegraphics[width=0.93\textwidth]{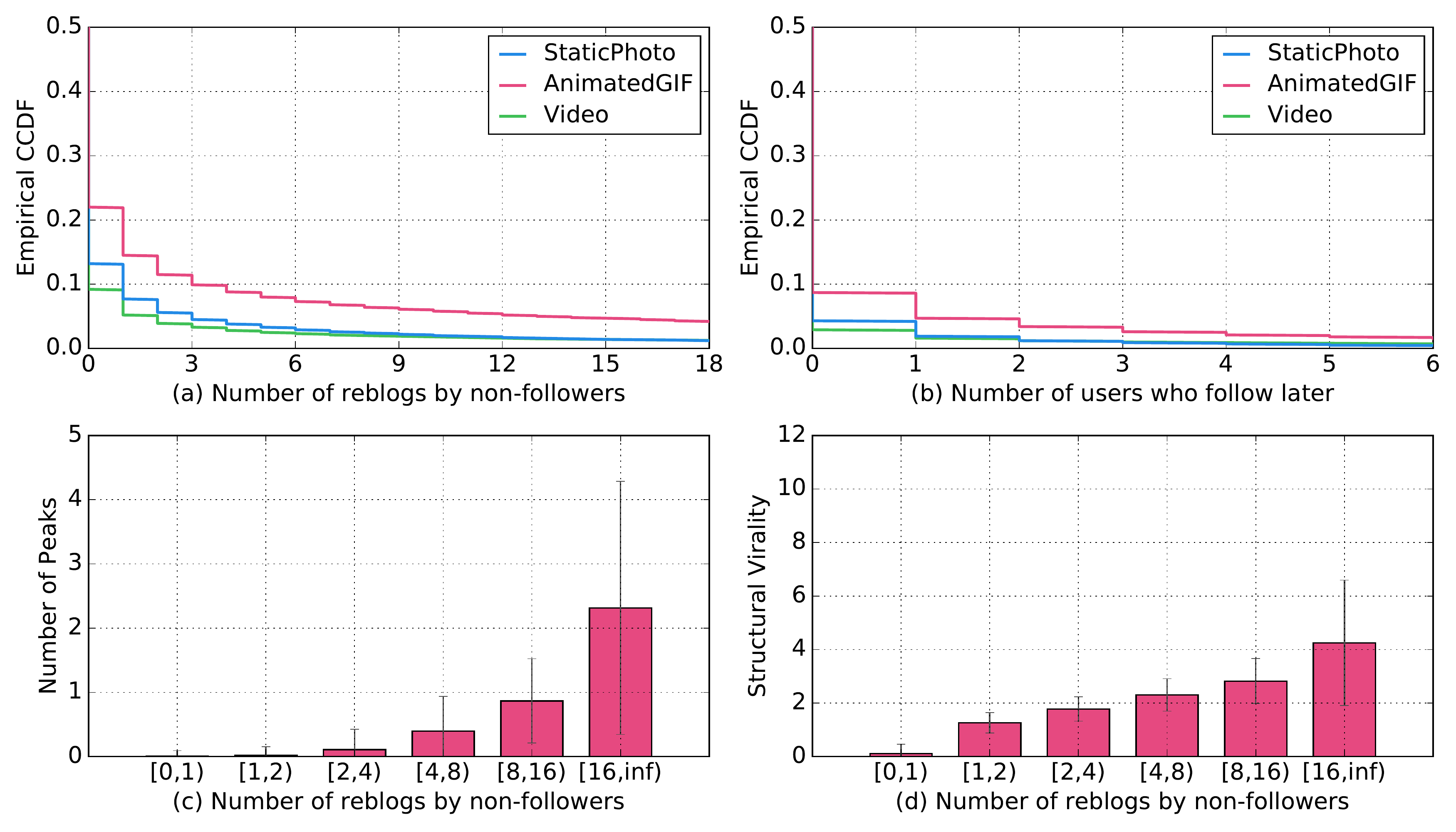}
   \vspace{-10pt}
  \caption{(a) CCDFs of the number of reblogs by non-followers, and (b) CCDFs of the number of users who follow later according to content types. (c)--(d) Correlation of the number of reblogs by non-followers (c) with the number of peaks, and (d) with the structural virality for animated GIF posts.
}
  \label{fig:outer_network}
\end{figure}

\smallskip
\textbf{What makes a post recur}?
Previously we found that the recurrence of a post is a strong indicator to a high structural virality. What causes the recurrence of a post then?

To analyze this, we check when a reblog occurs and when a child reblogger (\ie~a node in a cascade tree) follows its parent reblogger.
We count how many rebloggers in each cascade become the followers of their parents after reblogging a post.
That is, we count how many non-followers become followers after reblogging the followee's posts.

Figure~\ref{fig:outer_network} (a)-(b) summarize CCDFs of (a) the number of reblogs from non-followers and (b) the number of non-followers who later follow their parents after reblogging the posts. It shows that animated GIFs are reblogged 174.83\% more often by the non-followers than other content types, and followed 15.56\% more frequently by non-followers after they reblog the post. The two-sample KS tests with 10,000 random samples show that the differences between GIFs and others are statistically significant at $p<10^{-16}$. %
Our results show that including animated GIF on its post makes it more likely to spread out beyond the existing follower-followee network, and at the same time making the existing network richer.

Figure~\ref{fig:outer_network} (c) shows a strong positive correlation between the number of reblogs from non-followers and the number of peaks.
Also, as a result of it, Figure~\ref{fig:outer_network} (d) reveals that the number of reblogs from non-followers also is positively correlated with the structural virality. Consequently, the number of reblogs from the non-followers can be interpreted as an indicator to increase of the number of peaks and the structural virality.

These results may be used to predict whether a post would go viral or not, by monitoring the number of peaks or the number of reblogs from the non-followers. However, both pieces of information can only be retrieved long after a post is created, and thus real-time prediction is not possible.
Hence, in the next section, we carry out  a set of experiments to explore another way of virality prediction by content, to predict immediately after when a new post is created.

\begin{table}[t]%
  \footnotesize
  \centering
  \begin{tabular}{|p{0.08\textwidth}|p{0.18\textwidth}|p{0.05\textwidth}|p{0.58\textwidth}|}
    \hline
    \bf & \bf Name    & \bf Dim   & \bf Description   \\\hline
    \bf Image
          & ContentWidth      & \hfill1     & Image width. \\
    \bf features
          & ContentHeight     & \hfill1     & Image height. \\
          & Texture           & \hfill3     & Image texture by wavelet transformation on each channel. \\
          & RGBHistogram      & \hfill64    & Each R/G/B component into 4 bins at 5 FPS.\\
          & HSVHistogram      & \hfill64    & Each H/S/V component into 4 bins at 5 FPS. \\
          & \#Segments        & \hfill1     & Number of regions after mean-shift segmentation.\\
          & \#Objects         & \hfill5     & Likelihood of containing 0, 1, 2, 3, 4+ items, measured by~\cite{zhang-cvpr-15}.\\
          & ObjectEmbedding   & \hfill2048  & An image embedding vector from the \texttt{pool5} layer of ResNet-152~\cite{he-cvpr-16}.\\\hline
    \bf Video
          & FramesPerSecond   & \hfill1     & Number of frames in each second. \\
    \bf features
          & Duration          & \hfill1     & Duration of a video in second.\\
          & MeanFrameDiff     & \hfill1     & Mean frame difference of each video.\\
          & MotionEmbedding   & \hfill4096  & A motion embedding vector from the \texttt{fc6} layer of C3D~\cite{tran-iccv-15}. \\\hline
    \bf Text
          & \#Hashtags        & \hfill1     & Number of hashtags in each post. \\
    \bf features
          & CaptionLength     & \hfill1     & Length of text description in each post.\\
          & HashtagEmbedding  & \hfill100   & LDA~\cite{blei-jmlr-03} topic representation of hashtags trained for our dataset. \\
          & CaptionEmbedding  & \hfill100   & LDA~\cite{blei-jmlr-03} topic representation of text trained for our dataset. \\\hline
    \bf Social
          & \#Followers       & \hfill1     & Number of users that the post owner follows.\\
    \bf features
          & \#Followees       & \hfill1     & Number of users who follow the post owner.\\
          & $\Delta$Followers & \hfill1     & Number of users that the post owner follows during 7 months.\\
          & $\Delta$Followees & \hfill1     & Number of users who follow the post owner during 7 months.\\
          & $\Delta$Posts     & \hfill1     & Number of original posts that a user created during 7 months. \\
          & $\Delta$Reblogs   & \hfill1     & Number of original posts that a user created during 7 months. \\
          & \#Avg(Co-Followees) & \hfill1   & Average \# of users who also follow the post owner.\\\hline
    \end{tabular}
    \caption{A list of extracted features from Tumblr posts.\label{tab:list_of_features}}
\end{table}

\section{Virality Prediction by Content}
\label{sec:prediction_by_content}

Based on our quantitative measurement of virality in the previous section, we investigate what makes GIF viral by analyzing statistical relationship between virality scores and various features extracted from visual, textual, social cues.

It is worth mentioning that our dataset captures real-world user behavior and exhibits a long-tail distribution of virality. This, in turn, can contaminate statistical analysis results if done on the entire dataset, or uniformly randomly sampled dataset.
To make our analysis less affected by the skewed distribution, we perform stratified sampling on our dataset, enabling us to obtain a randomized samples of roughly 150,000 posts for each content type from six different ranges of the distribution (\eg 0, 1, (1,2], (2,3], (3,4], (4,5], 5+ for structural virality scores).
During sampling, we only consider the posts that contain at least one tag and one word with single downloadable content (\eg some posts may include multiple images and videos).
As a result, the resulting dataset for analyses in this section contains 300,000 posts (100,000 posts per content type).

\subsection{Feature Extraction}

We extract various features listed in Table~\ref{tab:list_of_features}. To compute image features from animated GIFs and videos, we sample videos by five frames per second up to one minute and average the features across sampled frames. For textual features, we lowercase the raw text, remove any HTML tags and replace each emoji to a single special character.

\subsubsection{Image Features}
\label{sec:image_features}$ $

\textbf{Content width and height}. We measure these from the resolution of visual content.

\textbf{Texture}. This feature encodes the graininess, smoothness, and directionality of an image. Based on \cite{arivazhagan-prl-03}, we apply to each color channel in the cylindrical color space (IHSL) of an image, the three-level Daubechies wavelet transform and average the output values.

\textbf{RGB and HSV histograms}. We use four bins per color channel. For GIFs and videos, we sample five frames per second from the first one minute and take an average of the histograms across frames.

\textbf{\#Segments and \#Objects}. These features encode the perceptual complexity of an image~\cite{totti-websci-14}. We count the number of segments using the mean shift segmentation algorithm~\cite{comaniciu-pami-02}, and the number of objects using the salient object subitizing algorithm~\cite{zhang-cvpr-15}.

\textbf{ObjectEmbedding}. We extract the \texttt{pool5} layer of ResNet-152~\cite{he-cvpr-16} pretrained on the ImageNet 2012 classification dataset~\cite{russakovsky-ijcv-15}.

\subsubsection{Video Features}
\label{sec:video_features}$ $

\textbf{FPS and Duration}. We obtain these features from the header file. The duration is measured in seconds.

\textbf{MeanFrameDiff}. We measure visual complexity of a video by computing the mean pairwise frame difference.

\textbf{MotionEmbedding}. We encode motion information in video using the C3D model~\cite{tran-iccv-15} pretrained on the Sports-1M dataset~\cite{karpathy-cvpr-14}. We use 16-frame input with a stride of 8, and take the max pool from the embedding at the \texttt{fc6} layer.

\subsubsection{Text Features}
\label{sec:text_features}$ $

\textbf{\#Hashtags and CaptionLength}. We count the number of hashtags and the number of characters in caption.

\textbf{HashtagEmbedding and CaptionEmbedding}. We encode semantic information from captions and hashtags using the LDA topic distribution~\cite{blei-jmlr-03}, training on one million hashtags and captions randomly sampled from our dataset.

\subsubsection{Social Features}
\label{sec:social_features}$ $

\textbf{\#Followers and \#Followees}. We count the number of users who the post owner follows (\#Followers) and who follow the post owner (\#Followees) on March 1, 2016.

\textbf{$\Delta$Followers and $\Delta$Followees}. We count the number of users who the post owner follows ($\Delta$Followers) and who follow the post owner ($\Delta$Followees) within March 1, 2016 and September 30, 2016.

\textbf{$\Delta$Posts and $\Delta$Reblogs}. We count the number of original posts and reblogs that the post owner has created during March 1, 2016 and September 30, 2016.

\textbf{Average \#Co-Followees}. We count the number of users who are followed by the post owner's followees on March 1, 2016 and divide it by \#Followees.

\subsection{Prediction Methods}
\label{sec:prediction_content}

There are different ways to evaluate the importance of individual features for explaining our target variable (structural variable scores). Similar to Cheng~\etal~\cite{cheng-www-16}, we do this with a binary classification problem; for a given post, we predict whether the structural virality of a cascade reaches the median of all the cascades or not. As exactly half the cascades have larger structural virality than the median by definition, random guessing achieves the accuracy of 50\%. Predicting virality with more effective features we extracted in the previous section would bring better result. %

For the evaluation, we use a linear SVM with an $\ell_2$-loss. We vary the penalty term $C = 10^{n}$, $n=\{-3:3\}$ and find the optimal setting with the grid search and five-fold cross validation. We perform 10-fold cross-validation for all input combinations and report the average accuracy over all ten results for the consistency of our results.

\begin{table}[t]%
  \small
  \centering
  \def\arraystretch{1.3}%
  \setlength{\tabcolsep}{4pt}
    \begin{tabular}{|l|c|c|c|c|c||c|}
    \hline
    Content     & Image & Text  & Video & Social & All        & Peak \\\hline
    StaticPhoto & 59.68 & 58.97 & N/A   & 63.48  & 65.89      & 82.71 \\
    AnimatedGIF & 58.75 & 65.01 & 57.58 & 63.38  & \bf 68.84  & \bf 85.46 \\
    Video       & 62.95 & 58.73 & 61.91 & 60.53  & 63.81      & 64.08 \\\hline
    \end{tabular}
    \caption{Mean accuracy rates of virality prediction obtained using a linear SVM classifier.\label{tab:mean_accuracy_of_prediction}}
\end{table}

\subsection{Results and Discussions}
\label{sec:eval_virality_class}

Table~\ref{tab:mean_accuracy_of_prediction} shows mean accuracy rates for different sets of features obtained using an SVM. We obtained the best results when we simply use  the information related to the first peak (the number of reblogs, duration of the bursts, height of the peak). This agrees with our argument in section~\ref{sec:virality}, but the peak information cannot be used in real time because it is measurable only after the first peak settles down. Thus, for the practicality, we exclude the peak information for the rest of our discussion. Except for the peaks, we obtain the mean accuracy rates of 68.84\% for animated GIFs, 65.89\% for static photos, and 63.81\% for videos. Next, we discuss some of the most interesting findings from our experiments.

\textbf{GIF virality is easier to predict than that of images and videos}. Table~\ref{tab:mean_accuracy_of_prediction} shows that the best virality prediction accuracy of GIF is 4.47\% relatively higher than images and 7.88\% relatively higher than videos. To see whether the differences are statistically significant, we performed paired sample t-tests comparing accuracy rates between GIFs \& images and GIFs \& videos.
We obtain better prediction results for GIFs than images ($p<0.01$) and videos ($p<10^{-13}$), which confirm that the differences are statistically significant.

\textbf{Text features are the most predictive of GIF virality.} Table~\ref{tab:mean_accuracy_of_prediction} shows that text features consistently provide the best performance in predicting GIF virality among different feature types. Paired sample t-tests comparing performances between different pairs of features (text \& image, text \& video, etc.) indicate that the differences are all statistically significant at $p<0.01$. %

We note that this result is somewhat contradictory to the previous findings~\cite{totti-websci-14,deza-cvpr-15} that showed social features play the most important role in predicting image virality. Although we observe similar results for image virality (see Table~\ref{tab:mean_accuracy_of_prediction}), textual features are the most dominant in predicting GIF virality. One possible explanation for it is that non-followers tend to look for GIFs by the text content, as discussed in section~\ref{sec:virality}, and thus such text searchability may be connected  to the virality.

\section{Conclusion}
\label{sec:conclusion}

We conducted a series of quantitative and comparative studies about the virality of animated GIFs, by analyzing over ten months of \textit{complete} posts logs on Tumblr. We reported on a number of interesting, new findings on related to animated GIFs: First, Animated GIFs are associated with 47.59\% shorter captions and 29.18\% more hashtags than images and videos. Second, Animated GIFs spread deeper by 23.62\% and wider by 208.27\% than other visual content. They also have a 10.47\% higher degree of \textit{structural virality} than the other post types. Third, animated GIFs receive 174.83\% more reblogs from the non-followers, and 15.56\% more follows from them afterward. Forth, animated GIFs lead the second peaks occur 19.42\% more with 3.24\% shorter lull period in between the first and its subsequent peaks. Fifth, they also have 31.97\% more number of recurrence with 5.52\% longer duration of the first reblog cycle. Lastly, GIF virality is 4.47\% more predictable than that of images and videos.

There are several future research directions that go beyond this work. First, we are interested in in-depth analyses and mathematical models that describe information diffusion of animated GIFs within the network. Second, since we mainly deal with content features using computer vision techniques for the prediction, it is demanding to extend the study to uncover the relation with other types of features, such as user-specific features. Finally, virality prediction is an intriguing topic to explore, although precise prediction is very challenging. From the industry perspective, automatic prediction of content virality can help increase the engagement of users to the social network service; also, it could bring important impacts on social media marketing, personalization, and recommendation.
These are the focus of our ongoing work.

\end{document}